\documentclass[twocolumn,showpacs,superscriptaddress,email,floatfix,longbibliography,prl]{revtex4-2}

\usepackage{header}
\usepackage{subfiles} 

\begin{document}
\title{Many-body radiative decay in strongly interacting Rydberg ensembles}
\author{Chris Nill}
\affiliation{Institut f\"ur Theoretische Physik, Universit\"at Tübingen, Auf der Morgenstelle 14, 72076 T\"ubingen, Germany}
\author{Kay Brandner}
\affiliation{School of Physics and Astronomy and Centre for the Mathematics and Theoretical Physics of Quantum Non-Equilibrium Systems, The University of Nottingham, Nottingham, NG7 2RD, United Kingdom}
\author{Beatriz Olmos}
\affiliation{Institut f\"ur Theoretische Physik, Universit\"at Tübingen, Auf der Morgenstelle 14, 72076 T\"ubingen, Germany}
\affiliation{School of Physics and Astronomy and Centre for the Mathematics and Theoretical Physics of Quantum Non-Equilibrium Systems, The University of Nottingham, Nottingham, NG7 2RD, United Kingdom}
\author{Federico Carollo}
\affiliation{Institut f\"ur Theoretische Physik, Universit\"at Tübingen, Auf der Morgenstelle 14, 72076 T\"ubingen, Germany}
\author{Igor Lesanovsky}
\affiliation{Institut f\"ur Theoretische Physik, Universit\"at Tübingen, Auf der Morgenstelle 14, 72076 T\"ubingen, Germany}
\affiliation{School of Physics and Astronomy and Centre for the Mathematics and Theoretical Physics of Quantum Non-Equilibrium Systems, The University of Nottingham, Nottingham, NG7 2RD, United Kingdom}

\begin{abstract}
When atoms are excited to high-lying Rydberg states they interact strongly with dipolar forces. The resulting state-dependent level shifts allow to study many-body systems displaying intriguing nonequilibrium phenomena, such as constrained spin systems, and are at the heart of numerous technological applications, e.g., in quantum simulation and computation platforms. Here, we show that these interactions have also a significant impact on dissipative effects caused by the inevitable coupling of Rydberg atoms to the surrounding electromagnetic field. We demonstrate that their presence modifies the frequency of the photons emitted from the Rydberg atoms, making it dependent on the local neighborhood of the emitting atom. Interactions among Rydberg atoms thus turn spontaneous emission into a many-body process which manifests, in a thermodynamically consistent Markovian setting, in the emergence of collective jump operators in the quantum master equation governing the dynamics. We discuss how this collective dissipation ---  stemming from a mechanism different from the much studied super- and sub-radiance --- accelerates decoherence and affects dissipative phase transitions in Rydberg ensembles.
\end{abstract}

\maketitle

\textit{Introduction --- } Rydberg gases allow to explore the interplay between strong interactions, external driving imposed by external fields and dissipation. This has led to a whole host of theoretical and experimental works, investigating, for example, dissipative phase transitions, the dynamics of epidemic spreading and critical phenomena \cite{malossi2014,marcuzzi2014,urvoy2015,letscher2017,gutierrez2017,helmrich2020}, as well as the dissipative preparation of correlated quantum states \cite{kraus2008,carr2013,hoenig2013,petrosyan2013,roghani2018}. Dissipation typically manifests through two processes, which are decoherence (of quantum superposition) and radiative decay \cite{loew2012,gaerttner2014,marcuzzi2014a,guardado-sanchez2018}. Decoherence leads to a gradual decay of quantum superposition that is formed between the high-lying Rydberg state and the atomic ground state from which the Rydberg state is excited. This process can be controlled by the phase-coherence of the excitation laser and by the temperature of the Rydberg gas. It is also influenced by strong interactions among Rydberg atoms \cite{letscher2017a,guardado-sanchez2018,signoles2021,schultzen2022}, which can be exploited for designing single photon absorbers and emitters \cite{bariani2012,dudin2012,honer2011,tresp2016,stiesdal2021}. Radiative decay, on the other hand, is an ubiquitous process which is caused by the coupling of the atomic dipole to the electromagnetic field. This results in the spontaneous emission of a photon from a Rydberg excited atom and a concomitant quantum jump from the Rydberg state to a low-lying electronic state, e.g., the ground state.

\begin{figure}
    \centering
    \includegraphics[width=.9\columnwidth]{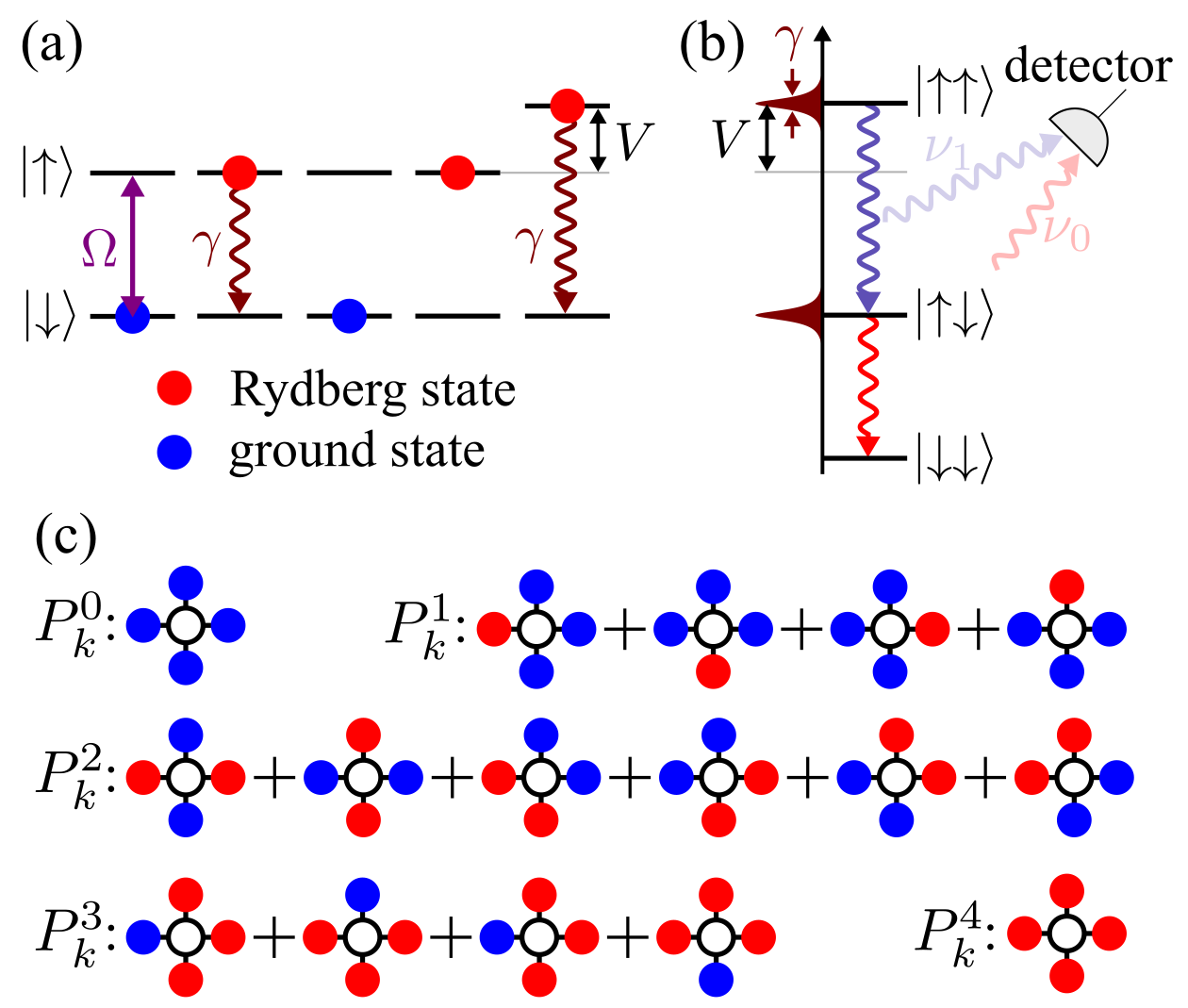}
    \caption{\textbf{Rydberg atoms and collective dissipation.} (a) One-dimensional lattice gas of interacting atoms resonantly driven by a laser with Rabi frequency $\Omega$. Neighboring atoms interact with interaction strength $V$ when simultaneously excited to their Rydberg state $\mid\uparrow\rangle$. Rydberg states decay under the emission of a photon to the ground state, $\mid\downarrow\rangle$, at rate $\gamma$. (b) Decay in a system of two atoms. When the interaction strength $V$ is larger than the natural linewidth $\gamma$ it is possible to discern whether a decaying Rydberg atom had an excited neighboring atom or not. This information can be inferred from the frequency of the emitted photon; $\nu_1$: excited neighboring atom, $\nu_0$: neighboring atom in the ground state. (c) Graphical representation of projectors $P^\xi_k$ which project on the subspace where the neighborhood of a reference atom (empty circle) contains $\xi$ excited atoms (in two dimensions). Due to the strong nearest-neighbor interaction an emitted photon carries information on the subspace from which the emission took place, leading to collective jump operators.}
    \label{fig:schematics}
\end{figure}
When considering ensembles of atoms, their coupling to the radiation field \cite{lehmberg1970,james1993} can lead to collective behavior as pointed out by Dicke in his seminal work \cite{dicke1954}. This  emerges when the typical distance between the atoms becomes comparable to the wavelength of the emitted radiation. In this case it is no longer possible to trace an emitted photon back to a specific atom. This loss of ‘which way’-information results in quantum interference that ultimately promotes this dissipation from a single-atom to a many-atom process. A striking consequence of this is the appearance of sub-radiant collective states whose lifetime may exceed that of single atoms by orders of magnitude \cite{asenjo2017,needham2019,rubies2022,zhang2020a,facchinetti2016,guerin2016,rui2020,ferioli2021}. In typical experiments, Rydberg atoms are separated by several micrometers. This is significantly larger than the wavelength for transitions to low-lying states, which is on the order of hundred nanometers. Radiative decay is therefore here not expected to acquire a collective character and is typically modelled as a single-atom process. Note, that super- and subradiance can nevertheless occur – and have been investigated – in Rydberg gases \cite{gross1979,wang2007,hao2021,suarez2022}. However, in these studies the considered radiative transitions take place among Rydberg states and the associated wavelengths are on the order of millimetres to centimetres \cite{saffman2010}.

In this work we demonstrate that strong interactions in Rydberg gases can nevertheless be responsible for another mechanism underlying collective dissipation. The fundamental observation is that the frequency of a photon that is spontaneously emitted from a decaying Rydberg atom depends on the state of the neighborhood of the emitting atom [cf.~Fig.~\ref{fig:schematics}(a-b)]. We unveil this effect and analyze its consequences in a simple setting, permitting for the exact derivation of the Markovian quantum master equation of the Rydberg gas which, as we discuss, features many-body jump operators. We show that the ensuing dissipation accelerates decoherence and that it further impacts on nonequilibrium phase transitions occurring in the stationary state of driven Rydberg gases. This collective {\it state-dependent} decay mechanism should be observable in (precision) experiments and is important for a thermodynamically consistent and faithful modelling of noise and error sources in quantum computers and simulators based on Rydberg atoms.

\textit{Interacting Rydberg gas in an electromagnetic field --- } To illustrate the above-mentioned effect we focus on a simple model of a Rydberg gas. The atoms are placed on the sites of a $d$-dimensional hypercubic lattice, labeled by the position vectors $\mathbf{r}_k$. Each atom is effectively described as a two-level (spin) system [see sketch in Fig. \ref{fig:schematics}(a)], with ground state $\mid\downarrow\rangle$ and  Rydberg state $\mid\uparrow\rangle$ separated by an energy difference $\omega_a$. We also assume for simplicity that the atoms only interact with their nearest neighbors with coupling strength $V$ [see Fig. \ref{fig:schematics}(a)]. This is accounted for by the Hamiltonian
\begin{equation}
H_\mathrm{atom}=\omega_a \sum_k n_k + \frac{V}{2}\sum_{|k-m|=1}n_k n_m, \label{eq:atomic_Hamiltonian}
\end{equation}
where $n_k=\mid\uparrow_k\rangle\! \langle\uparrow_k\mid$ is the projector on the Rydberg state of the atom located at position $\mathbf{r}_k$.

The atoms are immersed in an electromagnetic radiation field, described by the Hamiltonian
\begin{eqnarray}
H_\mathrm{rad}=\sum_{\mathbf{q},s} \omega_q a^\dagger_{\mathbf{q}s} a_{\mathbf{q}s}. \label{eq:photon_Hamiltonian}
\end{eqnarray}
Here, $a_{\mathbf{q}s}$ and $a^\dagger_{\mathbf{q}s}$ are the annihilation and creation operators of a photon mode with momentum $\mathbf{q}$, polarization $s$, and energy $\omega_q=c|\mathbf{q}|$ ($c$ is the speed of light). The dipole coupling between the atoms and the electromagnetic field modes is determined by the interaction Hamiltonian
\begin{eqnarray}
H_\mathrm{int}&=&\sum_{k,\mathbf{q},s} \left(g_{\mathbf{q}s} a^\dagger_{\mathbf{q}s} e^{i\mathbf{q}\cdot\mathbf{r}_k} +\mathrm{h.c.}\right)(\sigma_k^+ + \sigma_k^-). \label{eq:interaction_Hamiltonian}
\end{eqnarray}
Here, $\sigma_k^+=\left(\sigma_k^-\right)^\dagger=\mid\uparrow_k\rangle\! \langle\downarrow_k\mid$ is the atomic raising operator for the atom located at $\mathbf{r}_k$. The coupling constant $g_{\mathbf{q}s}=\sqrt{\frac{\omega_\mathbf{q}}{2\epsilon_0 {\cal V}}}(\mathbf{d}\cdot \bm{\varepsilon}_s)$ depends on the atomic transition dipole moment $\mathbf{d}$, the unit polarization vector $\bm{\varepsilon}_s$, the vacuum permittivity $\epsilon_0$ and the quantization volume ${\cal V}$. Note, that a variant of this model was considered also in Refs. \cite{lukyanets2006,lukyanets2011}, to study superradiance in the presence of interactions.

Our aim is to integrate out the electromagnetic field modes in order to obtain a quantum master equation that describes the open quantum dynamics of the atomic ensemble. We follow a procedure analogous to the usual one developed for the description of non-interacting atoms immersed in the radiation field (see e.g.~Refs.~\cite{lehmberg1970,james1993}). First, we rotate into the interaction picture with respect to the atom and radiation degrees of freedom via the unitary transformation $U=\exp\left[it (H_\mathrm{atom}+H_\mathrm{rad})\right]$. Due to the Rydberg interactions, atomic operators acquire an operator-valued phase which depends on the neighborhood of the considered atom, e.g.,
\begin{eqnarray}
U \sigma^+_k U^\dagger=\sigma^+_k \exp(i\omega_a t)  \exp(i V t \sum^{2d}_{\xi=0} \xi P^\xi_k).
\end{eqnarray}
Here, $\xi\,\in\,\{0,1, ..., 2d\}$ and $P^\xi_k$ is the projector on the subspace containing exactly $\xi$ excited atoms in the neighborhood of atom $k$  [see Fig.~\ref{fig:schematics}(c)]. Similar structures emerge, for example, in the so-called PXP model or the quantum hard-squares model, which describe strongly interacting Rydberg gases \cite{sun2008,lesanovsky2011}.

As derived in the Supplementary Material \cite{SM},\vphantom{\cite{breuer2002,redfield1957,Wilcox1967,qutip1,qutip2,cresser2010,levine2018}} after the Born-Markov and rotating-wave approximations, the quantum master equation reads (in the original lab frame):
\begin{align}
\dot{\rho}=-i[H_\mathrm{atom},\rho]+\gamma \sum_k \!\!\left[ \sum_{\xi=0}^{2d} P^\xi_k\sigma^-_k\rho \sigma^+_k P^\xi_k -\frac{1}{2}\{n_k,\rho\}\right],\label{eq:collective_master_equation}
\end{align}
where $\gamma=(|\mathbf{d}|^2 \omega^3_a)/(3\pi c^3 \epsilon_0)$ is the single atom decay rate. Note, that for the rotating-wave approximation to be valid the nearest-neighbor interaction strength $V$ must be much larger than $\gamma$. Moreover, we have assumed that $\omega_a\gg V$, which allows us to neglect corrections to the spontaneous emission rate originating from the interaction shift of the atomic levels in the presence of neighboring excitations, which are of order $\omega_a/V$. Finally, note that we consider the separation between neighboring atoms to be much larger than the transition wavelength $\lambda=2\pi c/\omega_a$. This allows us to neglect the effect of coherent dipole-dipole interactions and collective dissipation (i.e. super- and subradiance) induced by the radiation field. These conditions are typically met in current Rydberg quantum simulators using optical tweezer arrays. For example, in the experiment reported in Ref. \cite{scholl2021}, one finds for rubidium-$87$ atoms (principal quantum number $n=75$, lattice constant $10\,\mu$m, $C_6$-coefficient $C_6=\hbar\times 2\pi\times 1.947$ GHz $\mu$m$^6$): $\omega_a\approx 10^{15}$ Hz, $V=\hbar\times 2\pi\times 2\cdot 10^6$ Hz, $\gamma=6\cdot 10^3$ Hz.

From the master equation (\ref{eq:collective_master_equation}) one can read off that an atom at position $\mathbf{r}_k$ has $2d+1$ different decay channels, where $d$ is the dimension of the hypercubic lattice on which the atoms are positioned. Each of these channels, which is represented by the collective many-body jump operator $L_k^\mathrm{c} =\sqrt{\gamma} P^\xi_k \sigma_k^-$ corresponds to a different number of excited atoms $\xi$ in the atom's neighborhood, and can be associated to a different frequency of the emitted photons, $\nu_\xi=\omega_a + \xi V$  [see Fig. \ref{fig:schematics}(b)]. The many-body operators $L_k^\mathrm{c}$ are also consistent with thermodynamic considerations.
While we treat the background radiation field as an effective zero-temperature reservoir here, which is well justified since the atomic energy scale $\omega_a$ is typically much larger than the temperatures encountered in quantum-optical experiments, it is in principle straightforward to extend our approach to thermal environments with finite inverse temperature $\beta$.
One would then expect the Gibbs state 
$\propto\exp[-\beta(H_\text{atom})]$ to be a stationary state of the corresponding master equation.
This condition, which is indeed met in our many-body approach, is also both sufficient and necessary for consistency with the second law of thermodynamics, at least in situations where the standard weak-coupling, Born-Markov and rotating-wave approximations are applicable \cite{Spohn1978,Brandner2016}. On the other hand, a simpler model in which each atom would feature a single decay channel, represented by a jump operator $L_k^\mathrm{s} =\sqrt{\gamma} \sigma_k^-$ that does not account for interactions between atoms, would lead to a non-thermal stationary state at finite temperatures and thus, in general, to violations of the second law, see also Ref.~\cite{Levy2014}.

\textit{Decoherence dynamics --- } In order to analyze the impact of collective jump operators versus the conventionally employed  single-atom decay, we consider an atomic ensemble that is initially prepared in the state $| \Psi_0\rangle = (1/2)^{N/2}\bigotimes_k\left[\mid\downarrow\rangle_k+\mid\uparrow\rangle_k\right]$. Experimentally, such product state can be prepared in an interacting system by a strong laser pulse whose Rabi frequency $\Omega$ is much larger than the interaction strength $V$. We study the evolution of the average single-atom (Rydberg state - ground state) coherence, which can be measured experimentally \cite{semeghini2021}, and which we decompose as
\begin{align*}
    X(t)=\frac{1}{N} \sum_k \langle\sigma^-_k\rangle (t)
        = \frac{1}{N} \sum_{k,\xi} \langle P^\xi_k \sigma^-_k\rangle (t)
         = \sum_\xi X_\xi (t).
\end{align*}
The evolution equation of the expectation values $X_\xi$ is readily obtained \cite{SM}. For the collective dissipation, described by Eq.~(\ref{eq:collective_master_equation}), we obtain
\begin{eqnarray*}
\dot{X}^\mathrm{c}_\xi &=& -\left(i \omega_a +\frac{\gamma}{2}\right) X^\mathrm{c}_\xi - \xi\,\left(\gamma + i V\right) X^\mathrm{c}_\xi,
\end{eqnarray*}
while for the conventionally employed single-atom decay
\begin{eqnarray*}
\dot{X}^\mathrm{s}_\xi &=& -\left(i \omega_a +\frac{\gamma}{2}\right) X^\mathrm{s}_\xi - \xi\,\left(\gamma + i V\right) X^\mathrm{s}_\xi + \gamma \left(\xi+1\right) X^\mathrm{s}_{\xi+1}\, ,
\end{eqnarray*}
where we use the convention $X^\mathrm{s}_{2d+1}=0$. These equations can be exactly integrated with initial condition $X^{\mathrm{s}/\mathrm{c}}_\xi(0)=(1/N)\sum_k\langle \Psi_0 | P^\xi_k\sigma_k^-|\Psi_0\rangle=2^{-2d-1}\binom{2d}{\xi}$ \cite{SM}. Here, we focus on the short-time behavior, which already displays  a qualitative difference between collective dissipation and single-body decay:
\begin{eqnarray*}
|X^\mathrm{s}(t)| &\approx&  \frac{1}{2} - \frac{1}{4} \gamma t + \frac{\gamma^2-2d V^2}{16} t^2\, ,\\
|X^\mathrm{c}(t)| &\approx&  \frac{1}{2} - \frac{2d+1}{4} \gamma t + \frac{[(2d+1)^2+2d]\gamma^2-2d V^2}{16} t^2.
\end{eqnarray*}
For single-atom decay the initial drop of the coherence from its initial value $1/2$ is independent of the system geometry. The first collective contribution in $|X^\mathrm{s}(t)|$ emerges from the interaction of an atom with its neighbors, which involves the interaction strength $V$ and the coordination number $2d$ and is thus not of dissipative nature. In contrast, for the case of collective decay, already the leading term is dependent on the coordination number. This shows that collective dissipation notably accelerates the decoherence process as compared to the single-atom case. We briefly discuss the effect of collective dissipation on other coherence observables and on quantum correlations in \cite{SM}.

This effect should be even more dramatic in a continuous gas. Here, the initial rate of decoherence is proportional to the number of atoms, $N_\mathrm{int}$, with which a given reference atom interacts strongly enough so that the concomitant energy shift exceeds the single atom decay rate $\gamma$. For a homogeneous atomic gas with density $\varrho_0$ and Rydberg states that are interacting with a van-der-Waals potential \cite{saffman2010}, $V_\mathrm{vdW}(r)=C_6/r^6$, this number of atoms scales as $N_\mathrm{int}\sim\varrho_0 (|C_6|/\gamma)^{d/6}$ and thus the collective decoherence rate should scale as $\gamma_\mathrm{c}\sim \gamma \varrho_0 (C_6/\gamma)^{d/6}$.

\textit{Stationary state of a laser-driven Rydberg gas --- } The stationary state of the dynamics considered so far is the one devoid of any Rydberg excitation, since the system is only coupled with an effectively zero-temperature reservoir. In the following, we are interested in exploring the stationary state that emerges when (collective) radiative decay competes with external laser driving. To include the excitation laser (with frequency $\omega_l$, Rabi frequency $\Omega$ and detuning $\Delta=\omega_a-\omega_l$) we consider master equation (\ref{eq:collective_master_equation}) with the modified atomic Hamiltonian
\begin{eqnarray}
H_\mathrm{atom}\rightarrow H_\mathrm{atom} + \sum_k\left[\Omega \sigma^x_k + (\Delta-\omega_a) n_k\right]. \label{eq:laser_hamiltonian}
\end{eqnarray}
This is actually an \emph{ad-hoc} construction, given that the master equation has to be derived using the modified Hamiltonian. However, this approach is currently the standard one for incorporating coherent laser excitation, interaction and dissipation in interacting Rydberg gases \cite{loew2012,letscher2017,guardado-sanchez2018,morgado2021}. Our expectation at this point is that its analysis will reveal which quantitative and qualitative changes to the stationary state --- caused by collective jump operators --- one may expect.
\begin{figure}
    \centering
    \includegraphics[width=0.48\textwidth]{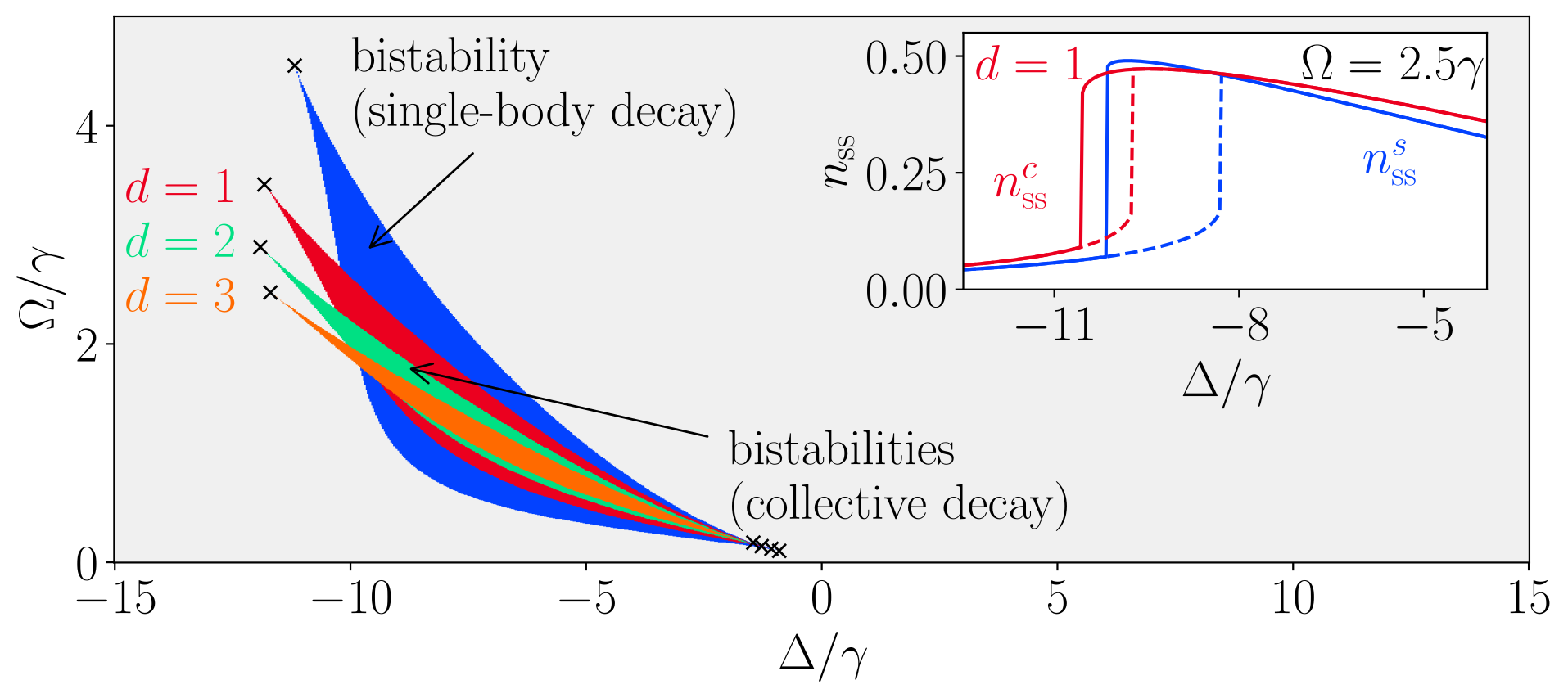}
    \caption{\textbf{Bistability region in the mean field phase diagram in the presence of single-atom and collective decay.} In the grey region the stationary state of the mean field equations is unique. In the colored region, whose shape depends on the dimension $d$, two stationary solutions exist. The cusps culminate in critical points, which are marked by crosses. The inset shows a cut of the stationary state density through the bistability region, taken at $\Omega=2.5\gamma$. The red (blue) curves show the stationary Rydberg excitation density $n_\mathrm{ss}^c$ ($n_\mathrm{ss}^s$) where the superscript $c$ ($s$) of collective (single-atom) decay. Solid and dashed lines mark the two stationary solutions. We have set $d\,V=10\gamma$. The mean field equations for $n_\mathrm{ss}^c$ are in Eq. \eqref{eq:mean-field} while those for $n_\mathrm{ss}^s$ are given in \cite{SM}.}
    \label{fig:mf_phase_diagram}
\end{figure}

We first perform a mean field analysis. Following the treatment of Ref. \cite{marcuzzi2014}, this leads to the mean field equations of motion
\begin{eqnarray}
	 \dot{n}        & = &   \Omega s_y  - \gamma n           \nonumber                                                         \\
	 \dot{s}_x  & = &-  \Delta s_y -\frac{\gamma}{2}(4 d\, n+1) s_x -2 d\,V n s_y \label{eq:mean-field}                   \\
	 \dot{s}_y &   = &- \Delta s_x -\frac{\gamma}{2}(4 d\, n+1)s_y  +2 d\,V n s_x -\Omega (4 n -2). \nonumber
\end{eqnarray}
Here, $n=\expval{n_k}$, $s_x=\expval{\sigma^k_x}$ and $s_y=\expval{\sigma^k_y}$ and translation invariance is assumed throughout. A noteworthy aspect of these equations is their dependence on the dimension $d$. In the contribution due to interactions, the latter enters through the combination $2d\,V$, with $2d$ being the coordination number of the $d$-dimensional hypercubic lattice. Therefore, different dimensions simply lead to a rescaling of the mean field interaction. This is not the case for the collective decay which results in terms proportional to $\gamma (4d\, n+1)$, which does not amount to a simple rescaling when changing dimensionality. This becomes visible in the stationary state phase diagram displayed in Fig. \ref{fig:mf_phase_diagram}. In the main figure we show the number of stable stationary mean field solutions. While for most parameters there is merely one solution, there exists a region for which two stationary solutions emerge. This bistability, extensively discussed in the literature, e.g. in Refs. \cite{lee2011,marcuzzi2014,letscher2017}, is seen in the inset. There we show the stationary Rydberg excitation density $n^c_\mathrm{ss}$ as obtained from the mean field equations (\ref{eq:mean-field}) after setting the time-derivatives to zero and solving for $n$. From a dynamical perspective bistability typically manifests in intermittency of the quantum jump statistics \cite{lee2012,ates2012,malossi2014}. The important aspect here is that the size and shape of this region strongly depend on the dimensionality, which is not the case when single-body decay is considered where one can simply rescale the interaction strength.

\begin{figure}
    \centering
    \hspace{-10pt} 
    \includegraphics[width=0.48\textwidth]{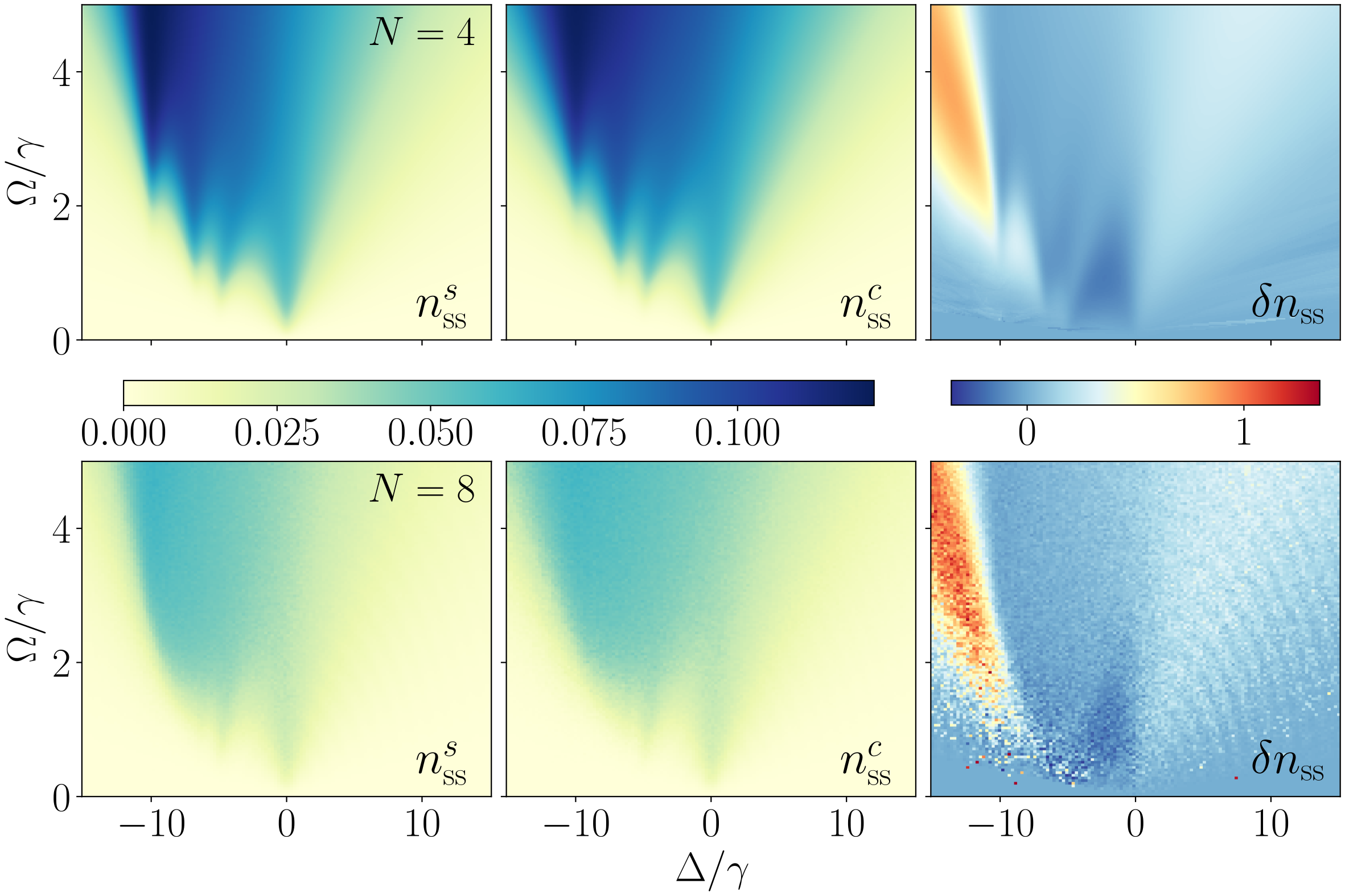}
    \caption{\textbf{Stationary state of a $d=1$ chain with periodic boundaries.} Shown is the stationary density of Rydberg excitations as a function of the laser detuning $\Delta$ and the Rabi frequency $\Omega$ with $V=10\gamma$ in the presence of single-body decay ($n^s_\mathrm{ss}$) and collective many-body decay ($n^c_\mathrm{ss}$). The data are obtained by solving Eq. (\ref{eq:collective_master_equation}) with Hamiltonian (\ref{eq:laser_hamiltonian}). In the rightmost column we show the relative difference between the two densities, $\delta n_\mathrm{ss}=(n_\mathrm{ss}^c - n_\mathrm{ss}^s)/n_\mathrm{ss}^s$. Significant deviations are visible for negative detunings in the region where bistable behavior is predicted by the mean field analysis. To estimate the steady state we average over $100$ linearly spaced data points in the interval $[4.75\, \gamma t,\, 5.00\,\gamma t]$. The oscillations visible in the region $\Delta /\gamma > 0$ are a finite time effect. The data for $N=4$ are obtained by exact diagonalization, while the $N=8$ data were calculated using continuous-time quantum jump Monte Carlo averaged over $300$ trajectories.}
    \label{fig:1d_simulations}
\end{figure}
To complement the mean field analysis we numerically calculate the stationary state of a small one-dimensional chain containing either $N=4$ or $N=8$ atoms. Qualitatively, both single-body and collective decay yield similar results, which are displayed in Fig.~\ref{fig:1d_simulations}. For negative detunings $\Delta$ --- where the mean field analysis predicts bistable behavior --- there is however a substantial quantitative difference. For example, the excitation density under collective dissipation can exceed the one predicted by single-atom decay by more than a factor two. This indicates that the bistable or metastable (in low dimensions) regions indeed are located in different regions of the parameter space, as the mean field result suggests. Note, that these features persist in the presence of weak laser phase noise \cite{SM}.

\textit{Conclusions and future directions --- } In this work we have studied the radiative decay of an interacting Rydberg gas. We have considered a rather simplified scenario, in which Rydberg atoms interact with nearest-neighbor interaction $V$, whose value exceeds that of the emission linewidth $\gamma$. Realistic interactions have a gradually decaying tail and there will be distances in which the interaction strength between the atoms becomes comparable with the decay rate. Here it is no longer possible to perform a rotating-wave approximation and the master equation becomes explicitly time dependent. Moreover, it would be interesting to include the laser driving systematically in the derivation of the master equation for the atomic system, e.g., by using the Floquet-Lindblad approach, which makes it possible to accommodate strong periodic driving fields in a thermodynamically consistent way \cite{Kosloff2013}.
For the sake of simplicity, we have focused on an ad-hoc approach in this article, where the driving is incorporated only in the unitary part of the master equation, as is currently standard quantum optics.

In order to experimentally probe the impact of collective effects it would be desirable to investigate strongly interacting Rydberg lattice systems that allow to observe dissipative dynamics over many emission cycles. This should be, for example, possible in trapped Rydberg ion systems \cite{schmidt-kaler2011,zhang2020}, which provide trapping of ground and Rydberg states alike and also offer the opportunity to continuously cool external degrees of freedom that may be heated from spontaneous emission.
\\
\textit{Data availability --- }Data is available through ancillary files of arXiv submission. Code can be downloaded from Zenodo: \url{https://doi.org/10.5281/zenodo.7400824}.

\acknowledgments
\textit{Acknowledgements --- }
The research leading to these results has received funding from the “Wissenschaftler-R\"uckkehrprogramm GSO/CZS” of the Carl-Zeiss-Stiftung and the German Scholars Organization e.V., through the Deutsche Forschungsgemeinsschaft (DFG, German Research Foundation) under Projects No. 435696605 and 449905436, as well as through the Research Unit FOR 5413/1, Grant No. 465199066. We also acknowledge support from the Baden-W\"urttemberg Stiftung through Project No.~BWST\_ISF2019-23. This work was supported by the University of Nottingham and the University of Tübingen's funding as part of the Excellence Strategy of the German Federal and State Governments, in close collaboration with the University of Nottingham.
K.B. acknowledges support from the University of Nottingham through a Nottingham Research Fellowship. This work was supported by the Medical Research Council [grant number MR/S034714/1; and the Engineering and Physical Sciences Research Council [grant number EP/V031201/1]. F.C.~is indebted to the Baden-W\"urttemberg Stiftung for the financial support by the Eliteprogramme for Postdocs.\\


\setcounter{equation}{0}
\setcounter{figure}{0}
\setcounter{table}{0}
\makeatletter
\renewcommand{\theequation}{S\arabic{equation}}
\renewcommand{\thefigure}{S\arabic{figure}}

\makeatletter
\renewcommand{\theequation}{S\arabic{equation}}
\renewcommand{\thefigure}{S\arabic{figure}}


\clearpage

\onecolumngrid
\setcounter{page}{1}

\begin{center}
{\Large SUPPLEMENTAL MATERIAL}
\end{center}
\begin{center}
\vspace{0.8cm}
{\Large Many-body radiative decay in strongly interacting Rydberg ensembles}
\end{center}
\begin{center}
Chris Nill$^{1}$, Kay Brandner$^{2}$, Beatriz Olmos$^{1,2}$, Federico Carollo$^1$ and Igor Lesanovsky$^{1,2}$
\end{center}
\begin{center}
$^1${\em Institut f\"ur Theoretische Physik, Universit\"at T\"ubingen,}\\
{\em Auf der Morgenstelle 14, 72076 T\"ubingen, Germany}\\
$^2${\em School of Physics and Astronomy and Centre for the Mathematics and Theoretical Physics of Quantum Non-Equilibrium Systems, The University of Nottingham, Nottingham, NG7 2RD, United Kingdom}
\end{center}

\section{I. Derivation of the master equation with collective jump operators}
In this section, we derive equation (5) from the main paper.
We start with the \textit{Redfield equation}, where the Born-Markov approximations have already been assumed \cite{breuer2002,redfield1957}. We label the trace over the bath degrees of freedom with $\tr{E}{\cdot}$ and the Hamiltonian in the interaction picture with $H''(t)$. The time evolution of the system at time $t$ is exclusively dependent on the current state $\rho_\mathrm{int}(t)=\vcentcolon\rho_\mathrm{int}$ in the rotating frame and reads:
\begin{equation}
	\dot{\rho}_\mathrm{int} =
	- \int_0^t \dd{\tau} \tr{E}{ \comm{H''(t)}{\comm{H''(\tau)}{\rho_\mathrm{int}\otimes \rho_E(0)}}}
	\label{eq:Redfield}\, .
\end{equation}
To obtain a Markovian master equation, the explicit dependence on the absolute time $t$ must be eliminated.
We therefore substitute $\tau=t-t'$ into \eqref{eq:Redfield} and assume that the bath correlation time $\tau_E$ is sufficiently small compared to $\tau$ $(\tau_E \ll \tau)$.
The integrand thus decays sufficiently fast, and the upper integration limit can be considered to be infinity to a good approximation. We now obtain the Markovian quantum master equation:
\begin{equation}
	\dot{\rho}_\mathrm{int} =
	-\int_0^\infty \dd{\tau}
	\tr{E}{\comm{H''(t)}{\comm{H''(t-\tau)}{\rho_\mathrm{int}\otimes \rho_E(0)}}}
	\label{eq:MarkovQMaster}\, .
\end{equation}
It should be noted that the above equation does not resolve the dynamics on time scales of the correlation time $\tau_\mathrm{E}$. In this sense, the equation describes a coarse-grained time evolution.
In the later calculation, we have to perform a rotating wave approximation (also called \textit{secular approximation}) which means that we have to neglect fast oscillating terms in the order of magnitude of the correlation time $\tau_\mathrm{E}$ by averaging.
Only in this way, we end up with a Markovian master equation that is a generator of a quantum dynamical semigroup \cite{breuer2002}.\\

The Hamiltonian of the whole system is obtained by summing up Eq. (1), (2) and (3) from the main paper:
\begin{align}
	H & = \ut{\omega_a\sum_k{n_k}
	+ \frac{V}{2} \sum_{\langle km \rangle }n_k n_m}{$H_\mathrm{atom}$}
	+ \ut{\sum_{\mathbf{q}s}{\omega_q a_{\mathbf{q}s}^\dagger a_{\mathbf{q}s}}}{$H_\mathrm{rad}$}
	+ \ut{\sum_{k,\mathbf{q},s} \left(g_{\mathbf{q}s} a^\dagger_{\mathbf{q}s} e^{i\mathbf{q}\cdot\mathbf{r}_k} +\mathrm{h.c.}\right)(\sigma_k^+ + \sigma_k^-)}{$H_\mathrm{int}$}
	\label{eq:HamiltonWithBath}\, .
\end{align}
We transform $H$ into the interaction picture via a unitary transformation:
\begin{align}
    H''(t)&=U H_\mathrm{int} U^\dagger\\
    U&=e^{it (H_\mathrm{atom}+H_\mathrm{rad})} \, .\label{eq:unitaryColl}
\end{align}
We then use the Baker-Campbell-Hausdorff formula \cite{Wilcox1967} to simplify $H''$ and subsequently obtain:
\begin{align}
	H''(t) & =\sum_{k,\mathbf{q},s}(g_{\mathbf{q}s} a_{\mathbf{q}s}^\dagger\e^{-i(\mathbf{q}\cdot \vb{r}_k-\omega_q t)} +\hc)
	\cdot \ut{(\sigma_k^+
	\e^{i t (\omega_a+ V\sum_{m \in \mathcal{I}_k}n_m)}
	+\hc)}{$\Chi_k(t)$}
	\label{eq:HInteractionPicture}\, .
\end{align}
We label all atoms in the neighborhood of atom $k$ with $m \in \mathcal{I}_k$.

Afterwards we introduce the projectors motivated and defined in figure (1) of the main paper.
This allows the exponential term in $\Chi_k(t)$ to be represented in the basis of the neighbouring atoms.
For $\Chi_k(t)$ one obtains:
\begin{equation}
	\Chi_k(t) =\sigma_k^+ \e^{it\omega_a} \sum_{\xi=0}^{2d} P_k^\xi \e^{i t \xi V} +\hc
	\label{eq:Chi_k}\, .
\end{equation}
We substitute $H''(t)$ from \eqref{eq:HInteractionPicture} into the Markovian quantum master equation \eqref{eq:MarkovQMaster}.
First, we simplify the occurring double commutator expression in \eqref{eq:MarkovQMaster}.
For the bath, we assume the vacuum state, i.e., a temperature $T=0$.
Thus, according to \cite{breuer2002}, we have for the expectation values of the bath operators:
\begin{align}
	\expval{a_\mathbf{q}}=\expval{a_\mathbf{q}^\dagger}                       =
	\expval{a_\mathbf{q} a_\mathbf{q'}}=\expval{a_\mathbf{q}^\dagger a_\mathbf{q'}^\dagger} =
	\expval{a_\mathbf{q}^\dagger a_\mathbf{q'}}                             =0
	 && \text{and}&& \expval{a_\mathbf{q} a_\mathbf{q'}^\dagger}                             =\delta_\mathbf{q,q'}\, .
\end{align}
After applying the trace over the degrees of freedom of the bath, the double commutator reads:
\begin{align}
	  & \tr{E}{\sum_{k,k'} \sum_{\mathbf{q},\mathbf{q'}} \sum_{s,s'} \comm{H''_{\mathbf{q}k}(t)}{\comm{H''_{\mathbf{q}'k'}(t-\tau)}{\rho_\mathrm{int}\otimes \rho_E(0)}}}\nonumber                 \\
	= & \sum_{k,k'}\sum_{\mathbf{q}s} |g_{\mathbf{q}s}|^2\e^{i\mathbf{q}\cdot  \vb{r}_{kk'}}\comm{\Chi_k(t)}{\Chi_{k'}(t-\tau) \rho_\mathrm{int}}\e^{-i \omega_q \tau} + \hc
	\label{eq:Chi-comm}\, .
\end{align}
Note that we now sum up pairwise over all atoms with index $k$ and $k'$.
If we add up all the intermediate results, the Markov master equation looks like this:
\begin{align*}
	\dot \rho_\mathrm{int} & = -  \sum_{k,k'}\sum_{\mathbf{q}s} |g_{\mathbf{q}s}|^2\e^{i\mathbf{q}\cdot  \vb{r}_{kk'}} \int_0^\infty \dd{\tau} \comm{\Chi_k(t)}{\Chi_{k'}(t-\tau) \rho_\mathrm{int}}\e^{-i \omega_q \tau} + \hc\,,
\end{align*}
where we have already interchanged integration and summation.

Next, we consider the integration of the time difference $\dd{\tau}$. To do this, we insert the expression for $\Chi_k(t)$ from \eqref{eq:Chi_k} and multiply the entire expression out. Summand by summand integrals of similar form are to be solved.
Using the formula from \cite{breuer2002}:
\begin{equation}
	\int_{0}^{\infty}  \,d\tau \e^{-i\tau(\omega_q\pm \omega_a)}=\pi \delta(\omega_q\pm \omega_a) - i \mathcal{P}\left(\frac{1}{\omega_q \pm \omega_a}\right)
\end{equation}
and application of the rotating wave approximation for all fast oscillating terms with factors $\mathcal{O}(\e^{i t V})$ or
$\mathcal{O}(\e^{2i t V})$, the equation can be further simplified.

Now the Markovian master equation is in a known form for several interacting atoms.
The summation over the bath modes $\mathbf{q}$ with polarisation $s$ is transformed into an integration analogously to  what is done in Ref.~\cite{james1993}.
After solving this one obtains a Markovian master equation:
\begin{equation}
	\dot\rho_\text{int} = \sum_{\xi=0}^{2d}\left(
	-i \sum_{k\neq k'}
	\tilde{V}_{kk'}^\xi \comm{\sigma_k^+ P_k^\xi P_{k'}^\xi \sigma_{k'}}{\rho_\mathrm{int}}
	+ \sum_{k,k'} \tilde{\Gamma}_{kk'}^\xi \left[\sigma_{k'} P_{k'}^\xi\rho_\mathrm{int} P_k^\xi  \sigma_k^+
		- \frac{1}{2}\acomm{\sigma_k^+ P_k^\xi P_{k'}^\xi \sigma_{k'}}{\rho_\mathrm{int}}\right]\right).
\end{equation}
Where
\begin{alignat}{3}
    \tilde{V}_{kk'}^\xi&=-&&\frac{3\gamma^\xi}{4}\left[ \alpha_{kk'} \frac{\cos{b_{kk'}}}{b_{kk'}}-\beta_{kk'} \left( \frac{\sin{b_{kk'}}}{b^2_{kk'}} + \frac{\cos{b_{kk'}}}{b^3_{kk'}} \right) \right]\, ,\\
    \tilde{\Gamma}_{kk'}^\xi&=&&\frac{3\gamma^\xi}{2}\left[ \alpha_{kk'} \frac{\sin{b_{kk'}}}{b_{kk'}}+\beta_{kk'} \left( \frac{\cos{b_{kk'}}}{b^2_{kk'}} - \frac{\sin{b_{kk'}}}{b^3_{kk'}} \right) \right]\, ,
\end{alignat}
utilizing $b_{kk'}=\omega_a \abs{\mathbf{r}_{kk'}}/c=2 \pi \abs{\mathbf{r}_{kk'}}/\lambda_a$, $\alpha_{kk'}=1-(\mathbf{d} \cdot \mathbf{r}_{kk'})^2$ , $\beta_{kk'}=1-3(\mathbf{d} \cdot \mathbf{r}_{kk'})^2$ and $\gamma^\xi= {\abs{\mathbf{d}^2}}(\omega_a+\xi V)^3/{3 \pi \epsilon_0 c^3}$.

We use now the assumptions of our model where $\omega_a \ll V$ thus $\gamma^\xi \approx \gamma$. Furthermore, we use the fact that we are in a regime where the distance $|\mathbf{r}_{kk'}|$ between the Rydberg atoms is large with respect to the wavelength of the atomic transition $\lambda_a$. Therefore the off-diagonal elements of $\tilde{V}_{kk'}$ and $\tilde{\Gamma}_{kk'}$ vanish. Moreover $\tilde{\Gamma}_{kk'} \approx \gamma \, \forall k=k'$.
Using the fact $\sum_{\xi} P_k^\xi =1$ to simplify the anti-commutator we obtain:
\begin{equation}
	\dot\rho_\text{int} =  \gamma \sum_k \left[ \sum_{\xi=0}^{2d}\sigma_k^- P_k^\xi\, \rho_\mathrm{int} \, P_k^\xi \sigma_k^+ -\frac{1}{2} \acomm{ n_k}{\rho_\mathrm{int}}\right]
	\label{eq:LindbladCollective}\, .
\end{equation}

Measurements and simulations are always carried out in the laboratory frame. For this reason, we transform back into the same laboratory frame.
It is valid using the chain rule and with $U$ from \eqref{eq:unitaryColl}:
\begin{align}
	\pdv{t} \rho
	=\pdv{t} \left( U^\dagger \rho_\text{int} U\right)
	=-i \comm{H_\mathrm{atom}}{\rho}
	+ U^\dagger \dot{\rho}_\text{int} U\, .
\end{align}
Let us now consider further the dissipative term. We apply Eq. \eqref{eq:LindbladCollective} and obtain:
\begin{align}
	U^\dagger \dot{\rho}_\text{int} U
	 & =  \gamma \sum_k \left[ \sum_{\xi=0}^{2d}
	\ut{U^\dagger \sigma_k^- P_k^\xi\, \rho_\text{int}\, P_k^\xi \sigma_k^+ U}{$\eqdef\mathcal{J}_1$}
	-\frac{1}{2} \ut{U^\dagger \acomm{n_k}{\rho_\text{int}}U}{$\eqdef\mathcal{J}_2$}\right] \, .
\end{align}
It can be easily shown by recalculation that:
\begin{equation}
	U^\dagger \sigma_k^- P_k^\xi U=e^{i\varphi} \sigma_k^- P_k^\xi
\end{equation}
holds, where $\varphi \in \mathbb{R}$.
Taking advantage of this identity, we transform the term $\mathcal{J}_1$:
\begin{alignat}{4}
	\mathcal{J}_1 & =U^\dagger      &  & \sigma_k^- P_k^\xi             &  & \rho_\text{int}             &  & P_k^\xi \sigma_k^+ U               \\[0.3em]
	              & =U^\dagger      &  & \sigma_k^- P_k^\xi U U^\dagger &  & \rho_\text{int} U U^\dagger &  & P_k^\xi \sigma_k^+ U               \\[0.3em]
	              & =e^{i \varphi}  &  & \sigma_k^- P_k^\xi             &  & \rho                        &  & P_k^\xi \sigma_k^+ e^{-i \varphi} \\[0.3em]
	              & =               &  & \sigma_k^- P_k^\xi             &  & \rho                        &  & P_k^\xi \sigma_k^+\, .
\end{alignat}
We find that the jump operators do not change due to the transformation into the laboratory frame.
For the transformation of $\mathcal{J}_2$, it follows analogously:
\begin{alignat}{3}
	\mathcal{J}_2 & =U^\dagger \acomm{n_k}{\rho_\text{int}}U
	              =\acomm{n_k}{\rho} \, .
\end{alignat}
It follows:
\begin{equation}
	\dot{\rho}= -i\comm{H_\mathrm{atom}}{\rho}+ \gamma  \sum_k \left[\sum_{\xi=0}^{2d}   \sigma_k^- P_k^\xi\, \rho\, P_k^\xi \sigma_k^+ -\frac{1}{2} \acomm{ n_k}{\rho}\right]
	\label{eq:MEQCollectiveLab}\, ,
\end{equation}
which is the master equation with collective jump operators in the laboratory frame Eq. (5) from the main paper.

\section{II. Decoherence dynamics}
We analyse the time evolution of the coherence $X(t)$ defined in Eq. (6) in the main text.
The time evolution can be calculated using the adjoint Lindblad operator $\mathcal{L}^\dagger$:
\begin{equation}
    \pdv{t}P^\xi_k\sigma^-_k=\mathcal{L}^\dagger [P^\xi_k\sigma^-_k]=\ut{i\comm{H_\mathrm{atom}}{P^\xi_k\sigma^-_k}}{coherent}+\ut{\mathcal{D}^\dagger[P^\xi_k\sigma^-_k]}{dissipative}\, .
    \label{eq:time_evolution_observable}
\end{equation}
For evaluating the coherent part we use
\begin{equation}
    i\comm{H_\mathrm{atom}}{P_k^\xi \sigma_k^-}=
    i\frac{V}{2} \sum_{l,m \in \mathcal{I}_l} \comm{n_{l} n_{m}}{P_k^\xi\sigma_k^-}
    +i\omega_a \sum_l \comm{n_l}{P_k^\xi \sigma_k^-},
\end{equation}
where we have labeled all atoms in the neighborhood of atom $l$ with $m \in \mathcal{I}_l$. By using the canonical commutator relations for Pauli matrices and the relations
\begin{align}
    \sum_{m \in \mathcal{I}_l} n_m &= \sum_{\eta=0}^{2d} \eta P_l^\eta \label{eq:projector_sum_identity}\\
    \comm{n_k}{P_l^\xi}&=0 \ \forall k,l\, 
\end{align}
one obtains
\begin{equation}
    i\comm{H_\mathrm{atom}}{P_k^\xi \sigma_k^-}=-i(\omega_a +\xi V)P_k^\xi \sigma_k^-.
\end{equation}
The dissipative part yields different results depending on whether single-atom or many-body decay is considered. For the collective many-body decay one finds:
\begin{equation}
    \mathcal{D^\dag_\mathrm{c}}[P_k^\xi \sigma_k^-]=\gamma \sum_{m} \left[
    \sum_{\eta=0}^{2d}P_m^\eta \sigma_m^+ P_k^\xi \sigma_k^- P_m^\eta \sigma_m^-
    -\frac{1}{2} \acomm{n_m}{P_k^\xi \sigma_k^-}
    \right]
    =
    -\gamma \left( \frac{1}{2}P_k^\xi \sigma_k^- + \sum_{m \in \mathcal{I}_k} n_m P_k^\xi \sigma_k^-\right).
\end{equation}
After applying Eq. \ref{eq:projector_sum_identity} this yields
\begin{equation}
\mathcal{D^\dag_\mathrm{c}}[P_k^\xi \sigma_k^-]=-\gamma \left( \frac{1}{2}+\xi \right) P_k^\xi \sigma_k^-.
\end{equation}
On the other hand, in the single-atom dissipation case, one obtains
\begin{align}
    \mathcal{D^\dag_\mathrm{s}}[P_k^\xi \sigma_k^-]
    &=\gamma \sum_m
    \left [
    \sigma_m^+ P_k^\xi \sigma_k^- \sigma_m^-
    -\frac{1}{2} \acomm{n_m}{P_k^\xi \sigma_k^-}
    \right]\\
    &=
    -\gamma\left(\frac{1}{2} P_k^\xi \sigma_k^- + \xi P_k^\xi \sigma_k^-
    -\sum_{m \in \mathcal{I}_k} \sigma_m^+ P_k^\xi \sigma_m^- \sigma_k^-\right).
\end{align}
Again, applying Eq. \ref{eq:projector_sum_identity} we find that
\begin{equation}
\mathcal{D^\dag_\mathrm{s}}[P_k^\xi \sigma_k^-]=-\gamma\left( \frac{1}{2} + \xi \right) P_k^\xi \sigma_k^- +\gamma(\xi+1) P_k^{\xi+1} \sigma_k^-.
\end{equation}
Taking the expectation value of Eq. (\ref{eq:time_evolution_observable}), summing over $k$ and dividing by the number of particles $N$ then yields for collective many-body decay:
\begin{eqnarray}
\frac{1}{N} \sum_k \expval{\pdv{t}P^\xi_k\sigma^-_k}=\pdv{t} \frac{1}{N} \sum_k \expval{P^\xi_k\sigma^-_k}=\pdv{t} X^c_\xi &=&  -i(\omega_a +\xi V) \frac{1}{N}\sum_k\expval{P_k^\xi \sigma_k^-}-\gamma \left( \frac{1}{2}+\xi \right)  \frac{1}{N}\sum_k \expval{P_k^\xi \sigma_k^-}\nonumber\\
&=&  -i(\omega_a +\xi V) X^c_\xi-\gamma \left( \frac{1}{2}+\xi \right)  X^c_\xi.
\end{eqnarray}
With single-body decay one obtains on the other hand
\begin{eqnarray}
\pdv{t} X^s_\xi &=& -i(\omega_a +\xi V) X^s_\xi -\gamma\left( \frac{1}{2} + \xi \right) X^s_\xi +\gamma(\xi+1) X^s_{\xi+1}.
\end{eqnarray}
These equations can be readily integrated. For the product state 
\begin{eqnarray}
| \Psi_0\rangle = \left(\frac{1}{2}\right)^\frac{N}{2}\bigotimes_k\left[\mid\downarrow\rangle_k+\mid\uparrow\rangle_k\right],
\end{eqnarray}
on obtains the initial values
\begin{eqnarray}
X^{\mathrm{s}/\mathrm{c}}_\xi(0)=\frac{1}{N}\sum_k\langle \Psi_0 | P^\xi_k\sigma_k^-|\Psi_0\rangle=2^{-2d-1}\binom{2d}{\xi}.
\end{eqnarray}
Using these, one finds (for $d=1,2,3$)
\begin{eqnarray}
|X^c(t)|=|\sum_\xi X_\xi^c(t)|=\frac{1}{2^{d+1}}e^{-\frac{(2d+1)\gamma t}{2}}\left[\cos(Vt)+\cosh(\gamma t)\right]^d
\end{eqnarray}
and
\begin{eqnarray}
|X^s(t)|=|\sum_\xi X_\xi^s(t)|=\frac{1}{2^{d+1}}e^{-\frac{(2d+1)\gamma t}{2}}\left[\frac{(\gamma^2+V^2)\cosh(\gamma t)+V\left(V\cos(Vt)+2\gamma\sin(Vt)\right)+2\gamma^2\sinh(\gamma t)}{\gamma^2+V^2}\right]^d,
\end{eqnarray}
whose short-time expansions are provided in the main text.

\section{III. Evolution of two-body correlations under many-body decay}
Here we briefly discuss the dynamics of two-body correlation functions and work out which difference arise between single-body and many-body dissipation. The effects are best observed in correlation functions of the type $\langle \sigma^+_k \sigma^-_m\rangle$, which involve coherences of the many-body state. For the sake of simplicity we consider a one-dimensional system, for which we obtain the following equations of motion for nearest and next-nearest neighbor correlations:
\begin{eqnarray}
\frac{\partial}{\partial t} \langle \sigma^+_k \sigma^-_{k+1}\rangle &=& iV \langle (n_{k-1}-n_{k+2}) \sigma^+_k \sigma^-_{k+1}\rangle \nonumber \\
&&-\gamma\left[\langle \sigma^+_k \sigma^-_{k+1}\rangle + \underline{\langle\left(n_{k-1}+n_{k+2}\right) \sigma^+_k \sigma^-_{k+1}\rangle}\right]\\
\frac{\partial}{\partial t} \langle \sigma^+_k \sigma^+_{k+1}\rangle &=& i2V\langle \sigma^+_k \sigma^+_{k+1}\rangle +iV \langle (n_{k-1}+n_{k+2}) \sigma^+_k \sigma^+_{k+1}\rangle \nonumber\\
&&-\gamma\left[\langle \sigma^+_k \sigma^+_{k+1}\rangle + \underline{\langle\left(n_{k-1}+n_{k+2}\right) \sigma^+_k \sigma^+_{k+1}\rangle}\right]\\
\frac{\partial}{\partial t} \langle \sigma^+_{k-1} \sigma^-_{k+1}\rangle &=& iV \langle (n_{k-2}-n_{k+2}) \sigma^+_{k-1} \sigma^-_{k+1}\rangle \nonumber\\
&&-\gamma\left[\langle \sigma^+_{k-1} \sigma^-_{k+1}\rangle + \underline{\langle\left(n_{k-2}+n_{k+2}\right) \sigma^+_{k-1} \sigma^-_{k+1}\rangle}\right]\\ 
\frac{\partial}{\partial t} \langle \sigma^+_{k-1} \sigma^+_{k+1}\rangle &=& iV\langle (n_{k-2}+2n_k+n_{k+2}) \sigma^+_{k-1} \sigma^+_{k+1}\rangle \nonumber\\
&&-\gamma\left[\langle \sigma^+_{k-1} \sigma^+_{k+1}\rangle + \underline{\langle\left(n_{k-2}+n_k+n_{k+2}\right) \sigma^+_{k-1} \sigma^+_{k+1}\rangle}\right]
\end{eqnarray}
In each of the equations, we have underlined the parts which are absent when the (conventional) single-body decay is considered. Single-body decay leads to the decay of correlations at a rate $\gamma$. Many-body decay leads in general to the emergence of density-dependent terms that further accelerate this decoherence, i.e. a density-dependent dephasing. However, not all correlation functions are equally affected. For example, in comparison to the next-nearest neighbor correlation $\langle \sigma^+_{k-1} \sigma^+_{k+1}\rangle$ the equation of motion for $\langle \sigma^+_{k-1} \sigma^-_{k+1}\rangle$ does not possess a term proportional to $n_k$ in the dissipative part. The reason for this difference is that the configurations $\mid\uparrow\rangle_{k-1} \mid\uparrow\!/\!\downarrow\rangle_k \mid\downarrow\rangle_{k+1}$ and $\mid\downarrow\rangle_{k-1} \mid\uparrow\!/\!\downarrow\rangle_k \mid\uparrow\rangle_{k+1}$ do not decohere relative to each other (due to many-body effects) since they are both located in the single excitation subspace, i.e., they are eigenstates (with eigenvalue $1$) of the projector $P^1_k$.\\
At the level of quantum trajectories, emission events associated with projectors with rank larger than one, i.e., all those projectors $P_{k}^\xi$ with $\xi\neq 0,2d$, can generate entanglement among spins which are in the neighborhood of the emitting one. This can be understood by focusing on the following simple case: lets consider a one-dimensional lattice with just three sites and take as initial state the one considered in the main text, $| \Psi_0\rangle = (1/2)^{3/2}\bigotimes_k\left[\mid\downarrow\rangle_k+\mid\uparrow\rangle_k\right]$. Imagine a trajectory in which the collective jump operator $P^1_2\sigma_2^-$ acts on this state. Then, after the emission occurred the state collapses onto $|\Psi_0'\rangle\propto P^1_2\sigma_2^-|\Psi_0'\rangle$, which is nothing but 
$$
|\Psi_0'\rangle=\frac{1}{\sqrt{2}}\left(\mid\uparrow\rangle_{1} \mid\downarrow\rangle_2 \mid\downarrow\rangle_{3} + \mid\downarrow\rangle_{1} \mid\downarrow\rangle_2 \mid\uparrow\rangle_{3}\right)\, .
$$
Thus, the neighbors of the emitting spin have been projected onto a Bell-pair state. This simple example, which can be generalized to jump operators associated with projectors onto any degenerate subspace, demonstrates that the collective jump operators can lead to entanglement in quantum trajectories. Nonetheless, the system (when not driven by a laser) converges to the state with all spins in $\mid\downarrow\rangle$ and thus entanglement is not possible asymptotically, not even in single trajectories.

\section{IV. Mean field equations}
In general, starting from a master equation in Lindblad form with operator $\mathcal{L}$ \cite{breuer2002}, one can derive a system of equations of motion for an observable $A$ with:
\begin{align}
	\expval{\pdv{t}A(t)} & =\expval{\mathcal{L}^\dagger A}\, .
\end{align}
In the analysis of our system, the average number of excitations in the system, normalized by the number of particles $N$, is of interest.
We define this quantity as the excitation density $\expval{n}$ of the system with:
\begin{equation}
	\expval{n}=\frac{1}{N}\sum_{k=1}^Nn_k \,.
\end{equation}
Based on the master equations we have obtained, for collective and single-atom dissipation, the equations of motion for $\expval{n_k}$ as well as for $\expval{\sigma_{x,y}^k}$ can be derived. Assuming translation-invariance in the system, i.e., $\expval{n_k}=\expval{n_{k'}}=n$ and $\expval{\sigma_{x,y}^k}=\expval{\sigma_{x,y}^{k'}}=s_{x,y}$ results in just three coupled equations:
\begin{alignat}{7}
	 & \dot{n}        &  & =  & \Omega s_y & - \gamma n                                                               &  &                                   & \nonumber                               \\
	 & \dot{s}_x &  & =- & \Delta s_y & -\frac{\gamma}{2}(\underline{\mathbin{4dn}}+1) s_x &  & -2 d V n s_y &                \\
	 & \dot{s}_y &  & =- & \Delta s_x & -\frac{\gamma}{2}(\underline{\mathbin{4d n}}+1)s_y  &  & +2 d V n s_x & -\Omega (4n - 2) \nonumber \,.
\end{alignat}
Only when using the collective jump operators, the term $4dn$ is present, which is underlined in the above equations.
In order to calculate the data in Fig. 2, we have put the left-hand side of these equations to zero and solved for the stationary density. This yields $n^c_\text{ss}$ when the underlined terms are included and $n^s_\text{ss}$ when the underlined terms are not included.

\section{V. Numerical computation of the Rydberg excitation density in the stationary state}
\begin{figure}[h]
    \centering
    \includegraphics[width=0.8\textwidth]{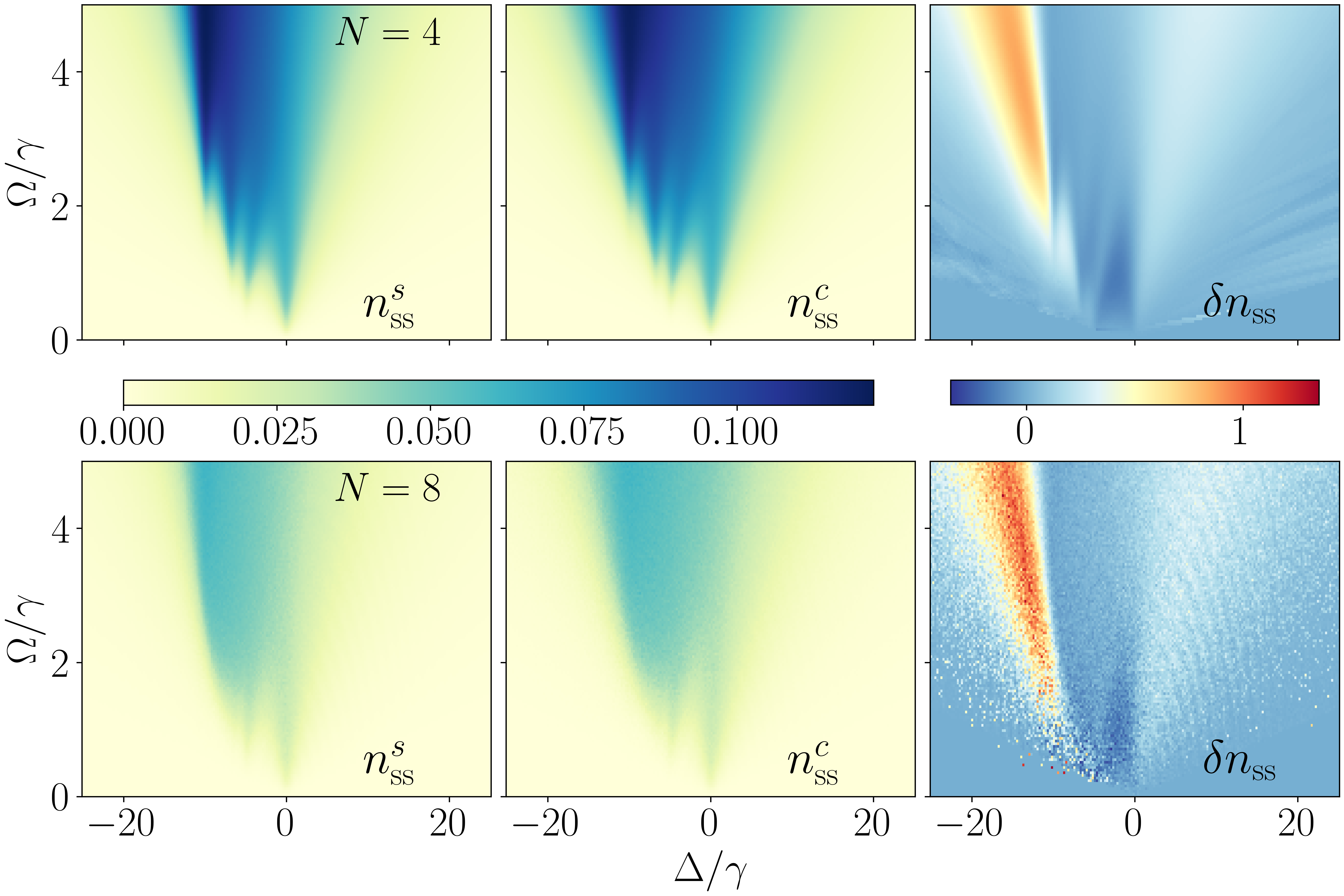}
    \caption{
    \textbf{Stationary state of a $d=1$ chain with periodic boundaries.}
   Shown is the stationary density of Rydberg excitations as a function of the laser detuning $\Delta$ and the Rabi frequency $\Omega$ with $V=10\gamma$ in the presence of single-body decay ($n^s_\mathrm{ss}$) and collective many-body decay ($n^c_\mathrm{ss}$).
   In the rightmost column we show the relative difference between the two densities, $\delta n_\mathrm{ss}=(n_\mathrm{ss}^c - n_\mathrm{ss}^s)/n_\mathrm{ss}^s$. Significant deviations are visible for negative detunings in the region where the mean field analysis predicts bistable behavior. In addition to Fig. 3 from the main text we extended the range of $\Delta/\gamma$ to $[-25,25]$.
   Further information on the simulation methods used can be found below.}
    \label{fig:1d_simulation_extended}
\end{figure}
To obtain the results of Fig. 3 of the main text  and of Fig. \ref{fig:1d_simulation_extended} we have solved Eq. (5) with the Hamiltonian given in Eq. (6) numerically.
To this end, we utilized the Python library QuTiP (version 4.7.0) \cite{qutip1,qutip2} which provides a powerful and easy-to-use framework for simulating open quantum systems.
For $N=4$ the dimension of the Lindblad generator is small enough that we can use exact-diagonalization methods to calculate the steady state.
This is done with the \textit{steady-state-solver} of QuTiP.
For $N=8$, due to performance reasons we instead simulate quantum jump trajectories exploiting the continuous-time Monte-Carlo algorithm \cite{saffman2010}.
In QuTiP this algorithm is implemented in the function \textit{mcsolve}.
To estimate the steady state we average over $100$ linearly spaced data points in the interval $[4.75\, \gamma t,\, 5.00\,\gamma t]$ for $3000$ trajectories.
The oscillations in Fig. \ref{fig:1d_simulation_extended} visible in the region $\Delta /\gamma > 0$ are a finite time effect.

\section{VI. The role of laser phase noise}
Throughout the paper, we have assumed that the atomic excitation is achieved through an ideal laser. Here, we briefly address the role of phase noise and how the latter affects our findings. Following, Ref. \cite{cresser2010} phase noise can be modelled through the dissipator
\begin{eqnarray}
    \mathcal{D_\mathrm{phase}}[\rho]= \sum_m \sum_{\alpha=x,y,z} \gamma_\alpha\left [\sigma_\alpha^m \rho \sigma_\alpha^m 
    -\rho \right],\label{eq:dissipator_phase_noise}
\end{eqnarray}
where the ``rates" $\gamma_\alpha$ are in general time-dependent. This dissipator, when acting alone, destroys  quantum coherence and leads to an infinite temperature, or completely mixed, state. This can be seen by inspecting the evolution equation of the operators (considering solely laser phase noise)
\begin{eqnarray}
\dot{n}_k\!\mid_\mathrm{phase} &=& \mathcal{D^\dagger_\mathrm{phase}}[n_k]=-\left(\gamma_x+\gamma_y\right)\left(2n_k-1\right)\\
\dot{\sigma}^x_k\!\mid_\mathrm{phase} &=& \mathcal{D^\dagger_\mathrm{phase}}[\sigma^x_k]=-2\left(\gamma_y+\gamma_z\right)\sigma^x_k\\
\dot{\sigma}^y_k\!\mid_\mathrm{phase} &=& \mathcal{D^\dagger_\mathrm{phase}}[\sigma^y_k]=-2\left(\gamma_x+\gamma_z\right)\sigma^y_k,
\end{eqnarray}
whose stationary state solution is $n_k=1/2$, $\sigma^x_k=\sigma^y_k=0$. In order to assess whether phase noise affects our findings in a qualitative fashion, we assume that the $\gamma_\alpha$ are constants and much smaller than the decay rate $\gamma$ of the Rydberg states. Only in this regime, i.e. with sufficiently weak phase noise, it is possible to speak about coherent laser excitations. Note, that in experiment the impact of phase noise can be controlled, e.g., through filtering, as shown in Ref. \cite{levine2018}.\\

\begin{figure}
    \centering
    \includegraphics[width=0.8\textwidth]{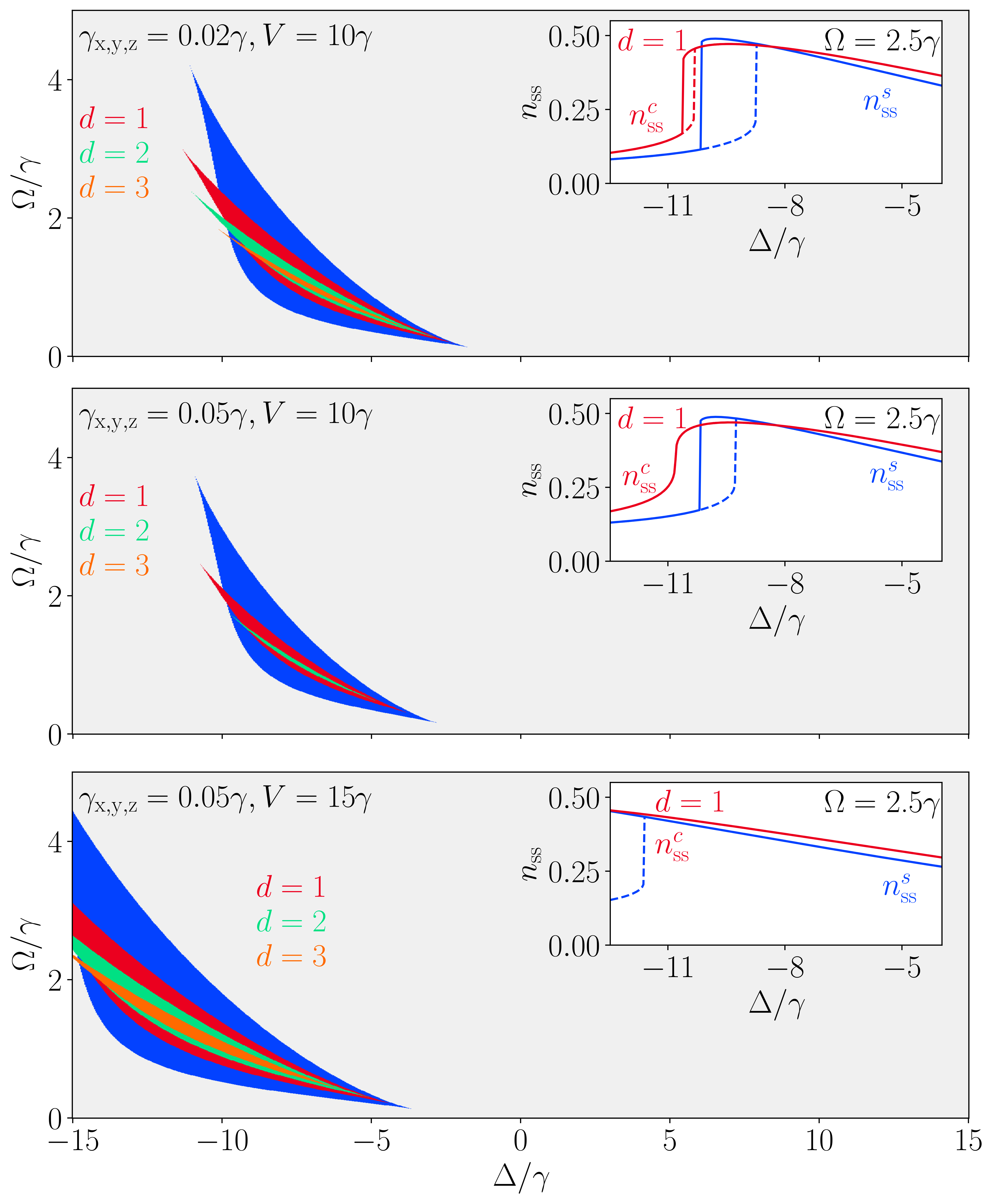}
    \caption{\textbf{Stationary state of a $d=1$ chain with periodic boundaries in the presence of laser phase noise.} Shown is the stationary density of Rydberg excitations as a function of the laser detuning $\Delta$ and the Rabi frequency $\Omega$ with $V=10\gamma$ in the presence of single-body decay ($n^s_\mathrm{ss}$) and collective many-body decay ($n^c_\mathrm{ss}$). From top to bottom we increase the strength of laser phase noise, which is modelled according to Eq. (\ref{eq:dissipator_phase_noise}). In the rightmost column we show the relative difference between the two densities, $\delta n_\mathrm{ss}=(n_\mathrm{ss}^c - n_\mathrm{ss}^s)/n_\mathrm{ss}^s$. Increasing the laser phase noise decreases the magnitude, i.e, the absolute values, of $\delta n_\mathrm{ss}$.}
    \label{fig:heatmap_noised}
\end{figure}

\begin{figure}
\centering
    \includegraphics[width=0.8\textwidth]{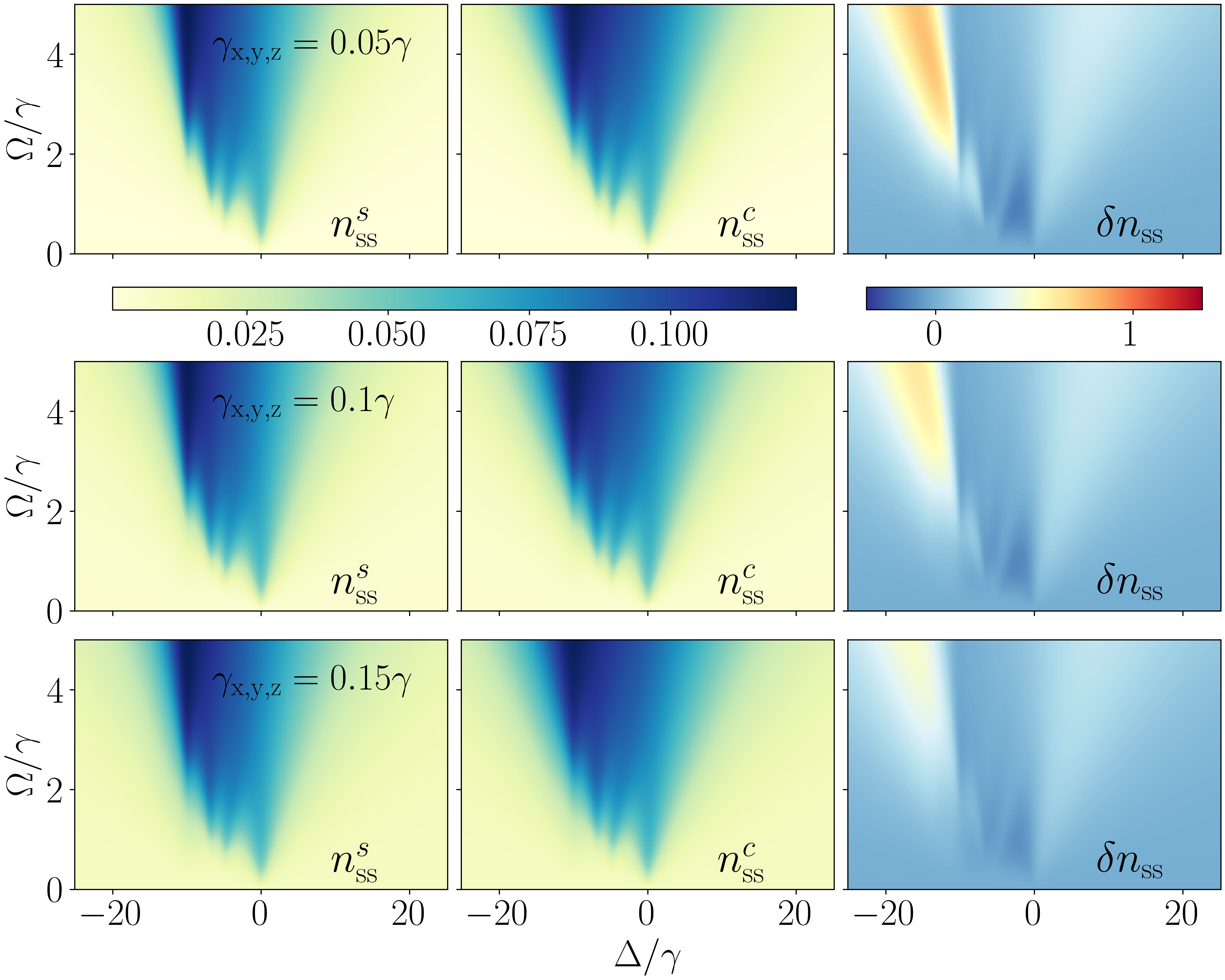}
    \caption{\textbf{Mean field phase diagram in the presence of dephasing noise.} When the rates $\gamma_\alpha$ [see Eq. (\ref{eq:dissipator_phase_noise})] are increased --- a situation which corresponds to an increase in laser phase noise --- the region, in the $\Delta-\Omega$-plane, in which bistability is observed shrinks (top to middle panel). This can be compensated by increasing the interaction strength $V$ (middle to bottom panel).}
    \label{fig:meanfield_noise}
\end{figure}

In Fig. \ref{fig:heatmap_noised} we show the same plot as in Fig. \ref{fig:1d_simulation_extended} but in the presence of phase noise. As can be seen, the deviations between the simulations with single-body and many-body decay become smaller as the strength of the phase noise increases. This is consistent since the larger the phase noise, the more the stationary state will tend to the infinite temperature state, no matter whether single-body or many-body decay is considered. Qualitatively similar behavior can be observed in the mean field phase diagram, which is shown in Fig. \ref{fig:meanfield_noise}. Phase noise decreases the size of the parameter region in which bistable solutions exist and can even lead to its complete removal. Indeed, in Ref. \cite{marcuzzi2014} it was shown that strong dephasing can remove the phase transition in dissipative Rydberg gases. Note, however, that bistable behavior has been observed experimentally \cite{malossi2014,letscher2017}. This corroborates our theoretical findings, that laser phase noise may lead to qualitative changes but does not destroy collective phenomena such as bistable behavior.




\end{document}